\documentclass[]{aa}
\usepackage{graphicx}

\begin{document}

\thesaurus{08.16.4, 08.03.4, 08.09.2 $\chi$~Cyg, 13.20.1}

\title{HCN in the inner envelope of $\chi$ Cygni}
\titlerunning{}
\authorrunning{Duari \& Hatchell}

\author{D. Duari\inst{1,2} \& J.~Hatchell\inst{1}}

\offprints{J.Hatchell@umist.ac.uk}

\institute{
   Department of Physics, UMIST, P O Box 88, Manchester M60 1QD,
   U.K. 
\and
   The Birla Institute of Astronomy and Planetary
Sciences, Calcutta 700 071, India
   }

\date{Received date / Accepted date}

\maketitle


\bigskip

\begin{abstract}

We have detected the $(0,1^{1c},0)\,J=3\hbox{--}2 $ and
$(0,0,0)\,J=8\hbox{--}7$ transitions of HCN towards the S star
$\chi$~Cygni.  The excitation requirements of these transitions are
too high to be satisfied in the outer envelope of the star, and the
emission must originate within $\sol 20$ stellar radii, ie. the
molecule must be forming close to the star.  This conclusion is
supported by a model for AGB stars in which molecules including HCN
form in a shocked wind close to the stellar surface.

%

\end{abstract}


\section{Introduction}

Recent detections of warm carbon bearing molecules in different O-rich
AGB stars suggest that these carbon species have to form in the deep
layers of the stellar wind.  Photochemical models (Willacy \& Millar
1997) of several O-rich stars, which rely on injection of certain
molecules to generate a carbon-rich chemistry at large radii, succeed
in reproducing the observed values of certain molecules but fail to
reproduce some molecular abundances, in particular that of HCN.  HCN
is observed in the envelope of several O- and S-type stars at
abundances of $>10^{-6}$ (Bieging \& Latter~1994 (BL94); Bujarrabal et
al. 1994; Sopka et al.~1989).  

Duari et al.~(1999) (DCW99) considered whether a shocked wind model
could reproduce HCN abundances for O-rich objects.  Local
thermodynamic equilibrium (LTE) models fail to reproduce the observed
abundances by several orders of magnitude (BL94).
An investigation of non-equilibrium chemistry of the oxygen rich Mira
IK Tau showed that these carbon bearing molecules can efficiently form
in the inner regions close to the stellar photosphere (DCW99). In this
case, it was shown that applications of stellar pulsation induced
shocks in a narrow region of the photosphere can give rise to
molecular processes in the immediate cooling layer and the
hydrodynamic cooling part ($=$~excursion) of the post shock region and
can produce certain carbon bearing molecules including HCN and CO$_2$.

It was earlier noted for S-type stars (Bujarrabal et al.~1994) that
their molecular line strengths are almost equidistant between those of
C- and O- rich objects. This property has been argued to be
related to the fact that the atmospheric C/O ratio is close to the
value of 1 and intermediate to that of O- and C- rich evolved stars.
We therefore apply DCW99's chemical model, which successfully
reproduces the observed HCN abundances in O-rich objects, to the
intermediate case of S stars.  In tandem we have looked for
observational evidence that the HCN is forming in the inner envelope
the S star $\chi$~Cyg.  Although HCN has previously been detected in
$\chi$~Cyg, the lines observed are dominated by material at large
radii, so here we look for high excitation transitions.  In carbon
stars, such transitions are detected in the inner regions of the winds
(eg. Lucas \& Guilloteau~1992).


In Sect.~\ref{section:observations} we describe the observations and
in Sect.~\ref{section:analysis} demonstrate that the HCN detected must
lie in the inner parts of the circumstellar envelope, a result which
is strongly supported by our modelling of the inner wind region of
$\chi$~Cyg (Sect.~\ref{section:model}).  Our conclusions are in
Sect.~\ref{section:conclusions}.

\section{Observations and results}
\label{section:observations}

Observations were made during 1999 using the James Clerk Maxwell
Telescope (JCMT) on Mauna Kea, Hawaii.  We observed the vibrationally
excited HCN~$(0,1^{1c},0)\,J=3\hbox{--}2$ line at 266~GHz, the
$(0,1^{1d},0)$ and $(0,2^{0},1)\,J=3\hbox{--}2$ lines at 267~GHz, and
HCN~$(0,0,0)\,J=8\hbox{--}7$ at 709~GHz.  The pointing centre was the
Hipparcos position of $\chi$~Cyg, $19^{\rm h}\, 50^{\rm m}\, 33.^{\rm
s}92\, {+32}\degr\, 54\arcmin\, 50.\arcsec6$ (Perryman et al.~1997).
Observing parameters are given in Table~\ref{tbl:obspars}.

\begin{figure}
\caption{HCN spectra towards $\chi$~Cyg: (top) HCN 8--7; (centre)
HCN~$(0,0,0)$ and $(0,1^{1c},0)\,J-3\hbox{--}2$ (also offset and
scaled by $1/100$); (bottom) H$^{13}$CN~3--2 with the positions of the
$(0,1^{1d},0)$ and $(0,2^{0},0)$ non-detections indicated. }
\begin{flushleft}
\includegraphics[angle=0, scale=0.85]{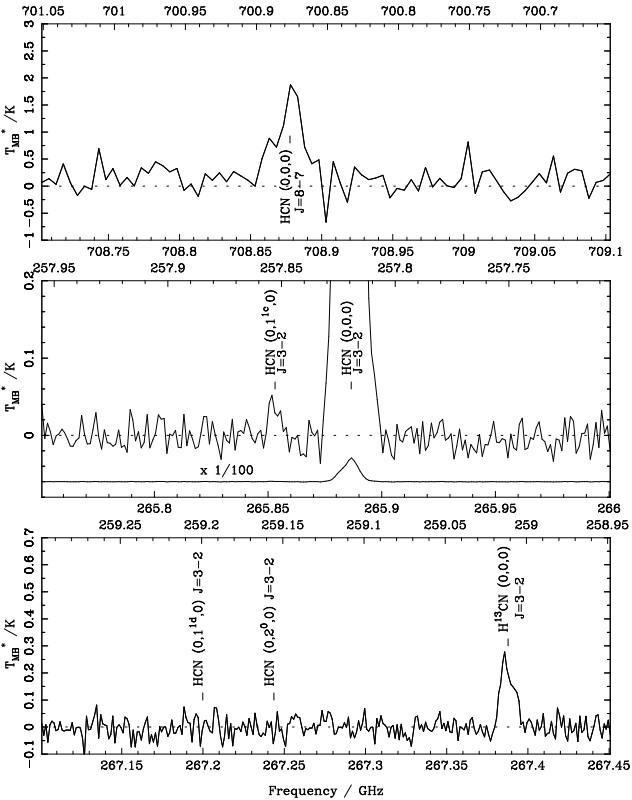}
\end{flushleft}
\label{fig:spectra}
\end{figure}

We detected HCN~$J=8\hbox{--}7$ and the $(0,1^{1c},0)\,J=3\hbox{--}2$
line at 266~GHz.  The spectra are shown in Fig.~\ref{fig:spectra}.
The HCN~$(0,1^{1d},0)$ and $(0,2^{0},0)$ lines at 267~GHz were not
detected: though this $v_2=1$ line should be the same strength as the
one detected at 266~GHz, this spectrum had a higher noise level (the
frequencies are marked in Fig.~\ref{fig:spectra}).  The
HCN and H$^{13}$CN~$J=3\hbox{--}2$ ground state lines at 266 and
259~GHz were also detected.

\begin{table}

   \caption[] {Observing parameters: transitions and frequencies;
   receivers; beam diameter (full width half maximum); zenith
   opacity at 225~GHz; main beam efficiency and system temperature;
   RMS noise levels on 1~km~s$^{-1}$ channels.}

   \begin{flushleft}
   \begin{tabular}{l c c c c}
    \hline
   \noalign{\smallskip}
line		&HCN 3--2 	&H$^{13}$CN 3--2      	&HCN 8--7\\
	&$(0,0,0)\&(0,1^{1c},0)$	&$(0,0,0)$       	&$(0,0,0)$\\
$\nu$/GHz	&265.8862 \& 	&259.0118	    &708.8772\\ 
		&265.8527	&      	&	\\
receiver	&A3		&A3            	&W\\
FWHM		&$18''$		&$18''$        	&$8''$\\
$\tau_{225}$  	&0.04		&0.1           	&0.04\\
$\eta_{\rm MB}$ &0.69		&0.69          	&0.30\\
$T_{\rm sys}$	&270		&400           	&3500--5000\\
$T_{\rm MB}$ (rms) &0.014	&0.030		&0.28\\
%

%

%
   \noalign{\smallskip}
   \hline
   \end{tabular}
   \end{flushleft}
 \label{tbl:obspars}
\end{table}

\section{Excitation analysis}
\label{section:analysis}

\begin{table*}

   \caption[] {Integrated intensities and upper state column densities for detected transitions.}
   \begin{flushleft}
   \begin{tabular}{l c c c c}
    \hline
   \noalign{\smallskip}
transition				&HCN 3--2 $v_2=0$		&H$^{13}$CN 3--2 $v_2=0$	&HCN 3--2 $v_2=1$		&HCN 8--7\\
frequency/GHz			&265.8862		&259.0118		&265.8527 		&708.8772\\ 
$\int T_{\rm MB}\,{\rm d}v/\hbox{K km s}^{-1}$	&$20.31$		&$1.38$			&$0.15$			&$0.30$\\		
$N_{\rm U}/\hbox{cm}^{-2}$			&$4.8\times 10^{12}$	&$3.4\times 10^{11}$	&$3.6\times 10^{10}$	&$1.1\times 10^{12}$\\
   \noalign{\smallskip}
   \hline
   \end{tabular}
   \end{flushleft}
 \label{tbl:results}
\end{table*}

The excitation requirements of the $J=8\hbox{--}7$ and
$(0,1^{1c},0)\, J=3\hbox{--}2$ transitions are high.  We consider
whether collisional or radiative excitation mechanisms can produce the
observed line intensities.

The column densities in each state can be determined from the observed
integrated line intensities, assuming they are optically thin.  For
HCN this conversion is:

\begin{equation}
\biggl({N_{\rm U}\over{\rm cm}^{-2}}\biggr) = 1.876\times 10^{13}
\biggl({\nu\over{\rm GHz}}\biggr)^{-1}{(2J+1)\over(J+1)}\biggl({\int
T_{\rm MB}\, {\rm d}v\over{\rm K km s}^{-1}}\biggr)
\label{eqn:columns}
\end{equation}
where $J$ is the rotational quantum number of the transition
$J+1\rightarrow J$, and we assume excitation temperatures much higher
than the 2.7~K background.  The resulting column densities are given
in Table~\ref{tbl:results}.

We use the H$^{13}$CN~$(0,0,0)\,J=3\hbox{--}2$ as a surrogate for the
equivalent H$^{12}$CN line as it is more likely to be optically thin.
We assume the same excitation in H$^{13}$CN as H$^{12}$CN, and a
[$^{12}$C]/[$^{13}$C] ratio, which is unknown in $\chi$~Cyg. Sopka et
al.~(1989) found [$^{12}$C]/[$^{13}$C]$>5$ from H$^{12}$CN/H$^{13}$CN~1--0 observations.
From the ratio of our H$^{12}$CN and H$^{13}$CN transitions,
[$^{12}$C]/[$^{13}$C]~$\sog 12$.  (This ratio also supports our
assumption that the H$^{13}$CN is optically thin.)  However, these low
limits do not rule out a much higher value.  We take as an upper limit
a typical ISM value of 70.

Assuming purely collisional excitation, 
\begin{equation}
{N_{\rm L}\over N_{\rm U}} = {g_{\rm L}\over g_{\rm U}} (n_{\rm crit}/n_{{\rm H}_2} + 1) e^{h\nu/kT_{\rm kin}},
\label{eqn:collisions}
\end{equation} 
where $N_{\rm U}$, $N_{\rm L}$, $g_{\rm U}$ and $g_{\rm L}$ are the
upper and lower state column densities and degeneracies and $n_{{\rm
H}_2}$ the unknown hydrogen column density.  The critical densities
for both transitions are high: for the $(0,1^{1c},0)\,J=3$ state we
calculate $n_{\rm crit} \simeq 4.8 \times 10^{11}\hbox{ cm}^{-3}$
(following Stutzki et al.~1988), and for $J=8\hbox{--}7$ $n_{\rm crit}
\simeq 1.2\times 10^9 \hbox{cm}^{-3}$.  Applying
Equation~\ref{eqn:collisions} together with the column densities from
Table~\ref{tbl:results}, the minimum density requirements to achieve
the observed ratios are $1\times 10^9$ and $5\times 10^7\hbox{
cm}^{-3}$ for $(0,1^{1c},0)\,J=3\hbox{--}2$ and $J=8\hbox{--}7$
transitions respectively.  For $J=8\hbox{--}7$ we assume the
population in $J=7$ is the same or less than that measured in $J=3$.

The envelope gas density falls off very rapidly with radius
(Bertschinger \& Chevalier~1985; Cherchneff et al.~1992).  Assuming
the stellar parameters for $\chi$~Cyg given in Table~\ref{tab:params},
the total hydrogen density at $1 R_{\star}$ is $3.6\times
10^{15}\hbox{ cm}^{-3}$ and the density falls below the density
requirement for $(0,1^{1c},0)\,J=3\hbox{--}2$ within $3.5 R_{\star}$ and
for $J=8\hbox{--}7$ within $5 R_{\star}$.  Alternatively,
assuming a density profile due to steady mass loss of $1.8\times 10^{-7}
\hbox{ M}_{\sun} \hbox{ yr}^{-1}$ at $8.9\hbox{ km s}^{-1}$ (BL94),
then the density at $1 R_{\star}$ is $7.5\times 10^8\hbox{ cm}^{-3}$
and falls as $R^{-2}$.  This constrains the HCN~8--7 emission to
within $\sim 4 R_{\star}$, assuming it is collisionally excited.

An alternative means of exciting these HCN transitions is by radiative
pumping by the IR radiation from the star.  The $(0,0,0)$ and
$(0,1,0)$ states are connected by $14\mu$m radiation with an Einstein
A coefficient of $3.7\hbox{ s}^{-1}$ (Ziurys~1986).  If collisions can
be neglected, and assuming geometrical dilution of the stellar
radiation field, the ratio of column densities in the ground and
$v_2=1$ vibrational states depends on the distance
from the star according to the formula:

\begin{equation}
{N_{\rm L}\over N_{\rm U}} = (e^{h\nu/k
T_{\star}}-1)(2R/R_{\star})^2 + 1
\label{eqn:columnratio}
\end{equation}
with $R$ radius and $T_{\star}$ stellar temperature.  Assuming a
stellar temperature of 2200~K, this constrains the radius at which the
$(0,1^{1c},0)$ line is emitted to $R \sol 17 R_{\star}$.

The $(0,0,0)\,J=8$ state cannot be directly pumped by radiation (as $\Delta
J = 0,\pm 1$ for HCN) but it could be populated through successive
excitations to the $(0,1,0)$ levels in increasing $J$ states.  The radius
requirements for this to take place are stronger than for the
$(0,1^{1c},0)$ state.

For either collisional or radiative excitation, the radius
requirements are tightened further if either (a)
[$^{12}$C]/[$^{13}$C]$ <70$ or (b) H$^{13}$CN~$J=3\hbox{--}2$ emission
originates from a larger region than HCN~$(0,1^{1c},0)\,J=3\hbox{--}2$
or HCN~$J=8\hbox{--}7$.

To conclude, the excitation requirements are such that both the
HCN~$(0,1^{1c},0)\,J=3\hbox{--}2$ and HCN~$J=8\hbox{--}7$ transitions
must originate from gas within $\sol 20$ stellar radii, in the inner
envelope of the star, whether collisionally or radiatively excited.

\section{Model of the inner wind of $\chi$ Cyg}
\label{section:model}

We have used our chemical model, which has succesfully produced the
observed HCN abundance in the case of an O-rich object (see DCW99 for
details), with a value of C/0 $\sim$ 1 for modelling the inner wind
region of $\chi$ Cyg.  The stellar parameters considered in this study
are listed in Table~\ref{tab:params}. The radius is obtained from the
observational value of Tuthill et al.~(1999) and the pulsation period is
from Bedding \& Zijlstra (1998). Using the standard pulsation equation
for Miras of Fox and Wood~(1982) and assuming that the star is
pulsating in its fundamental mode (Q =0.09), we derive a stellar mass
of 1.15 M$_\odot$, which is a typical value for the stars of this
class.  The temperature of 2200 K was obtained from Haniff et al.~(1995) 
which gave a luminosity which is close to the canonical value for
typical Miras.

We have considered here the inner wind above the photospheric region 
which experiences passage of strong, periodic shocks generated by stellar 
pulsation. The model deals with the chemistry of the immediate region 
(thermal cooling region) and the hydrodynamical cooling region of post 
shock described by Bertschinger \& Chevalier~(1985), Fox \&  Wood (1985) 
and Willacy \& Cherchneff (1998). 
\begin{table}
\begin{flushleft}
\caption{ $\chi$~Cyg - stellar parameters:
$\gamma$ and $\alpha$ are  defined as in Cherchneff et al. (1992) and
X (Y) $\equiv {\rm X \times 10^Y}$. }
\begin{tabular}{llll}
\hline
D	     & 106~pc  &$\dot {\rm M}$	&$1.8(-7)\hbox{ M}_\odot
\hbox{ yr}^{-1}$\\
T$_{\star}$  & 2200 K  & L$_\star$ & 1.8(3) L$_\odot$\\
R$_\star$    & 290 R$_\odot$  & M$_\star$ & 1.15 M$_\odot$\\
P            & 408 days  & n(r$_{\rm shock}$)  & 3.6 (15) cm$^{-3}$\\
$r_{\rm shock}$        & 1.0 R$_\star$   & $\alpha$  & 0.6\\
$\gamma =v_{\rm shock}/v_{\rm esc}$ & 0.89 & C/O ratio & 0.95 \\
\hline
\end{tabular}
\end{flushleft}
\label{tab:params}
\end{table}

\begin{table*}
\begin{flushleft}

\caption{Pre-shock, shock front and excursion
($\equiv$ post-shock) gas temperature and number density as a function
of position in the envelope and shock strengths. M is the Mach number
associated with each shock speed.}

\begin{tabular}{ccccccccc}
\hline
Position & Shock Vel. & M &\multicolumn{2}{c}{Pre-shock} & \multicolumn{2}{c}{
Shock Front} & \multicolumn{2}{c}{Start of excursion}  \\
 ($R_\star$) & (km s$^{-1}$) &{\rm } &$T_0$ (K) & $n_0$ (cm$^{-3}$) &
$T$ (K) & $n$ (cm$^{-3}$) & $T$ (K) & $n$ (cm$^{-3}$) \\
\hline
1.0 &32.0 &  10.3 & 2200 &3.62 (15) &
47458 & 2.07 (16) & 6528 & 1.26 (17) \\
1.5 & 26.1 &  9.51 &1724 & 1.28 (13) &
31959 & 7.28 (13) & 4867 & 3.98 (14) \\
2.0 & 22.6 & 8.98 & 1452 & 3.93 (11) &
24128  & 2.22 (12) & 3966 & 1.13 (13) \\
\hline
\end{tabular}
\end{flushleft}
\label{tab:jump}
\end{table*}

   \begin{table}
      \caption{Calculated fractional abundances (relative to the total
gas number density) versus shock strength and radius. 
}
\label{tbl:abundances}
\begin{flushleft}
\begin{tabular}{l l l l  l}
\hline\noalign{\smallskip}
 Species  &   T.E.  & 32 km/s        & 26.1  km/s        & 22.6
km/s        \\
  & 1. R$_{\star}$ & 1. R$_{\star}$ &  1.5 R$_{\star}$ & 2. R$_{\star}$
 \\
            \noalign{\smallskip}
            \hline
            \noalign{\smallskip}

H         &  1.0(-01)   &  4.3(-03)   &  4.4(-03)   &  4.2(-01)      \\
H$_2$        &  7.3(-01)   &  8.2(-01)   &  8.1(-01)   &  4.4(-01)      \\
S         &  1.5(-05)   &  2.0(-06)   &  2.9(-06)   &  1.6(-05)      \\
C$_2$H$_2$&  1.5(-15)   &  3.6(-11)   &  4.6(-12)   &  4.9(-15)      \\
CS        &  3.4(-09)   &  1.5(-05)   &  2.0(-05)   &  3.0(-06)      \\
HS        &  4.8(-06)   &  4.3(-06)   &  1.1(-06)   &  7.0(-09)      \\
H$_2$S       &  4.8(-07)   &  4.8(-07)   &  4.8(-08)   &  7.2(-13)      \\
NS        &  4.6(-11)   &  7.3(-12)   &  2.9(-11)   &  9.4(-13)      \\
N         &  3.4(-09)   &  3.0(-13)   &  7.6(-17)   &  1.1(-17)      \\
N$_2$        &  9.0(-05)   &  5.8(-06)   &  5.3(-05)   &  7.1(-05)      \\
{\bf HCN}$^\dagger$ & {\bf 1.6(-09)} & {\bf 1.8(-04)} & {\bf 8.5(-05)} &
{\bf 4.8(-06)}      \\
CN        &  1.4(-11)   &  4.6(-09)   &  6.2(-10)   &  1.9(-09)       \\
O         &  2.6(-08)   &  8.4(-11)   &  1.1(-11)   &  8.7(-10)       \\
OH        &  2.0(-07)   &  3.0(-08)   &  3.7(-09)   &  1.1(-08)       \\
H$_2$O       &  7.4(-06)   &  2.0(-04)   &  1.1(-04)   &  5.6(-06)       \\
CO        &  9.9(-04)   &  8.5(-04)   &  8.6(-03)   &  6.6(-03)       \\
CO$_2$       &  2.0(-09)   &  6.4(-08)   &  8.3(-07)   &  4.7(-06)       \\
O$_2$        &  1.2(-13)   &  6.8(-15)   &  5.5(-16)   &  5.3(-15)       \\
SiO       &  4.4(-05)   &  4.7(-05)   &  5.0(-05)   &  3.9(-05)       \\
            \noalign{\smallskip}
            \hline
\end{tabular}
{$^\dagger$ \it The observation value for HCN (averaged over the entire
envelope) is {\bf 2.5(-6)} derived from millimeter observations
(Bujarrabal et al.~ 1994)}
\end{flushleft}
\end{table}

\subsection{The chemistry}

We have considered 760 reactions involving 68 chemical species and have 
all possible chemical routes possible in a dense gas. 
The reaction rates 
considered are the same as in DCW99 (and the references therein). 
We have assumed thermal equlibrium for the photosphere and derived molecular 
abundances for the temperature, C/O ratio and gas density given in Table 3. 
We then apply a shock to the photosphere (the shock velocity considered 
is $v_{shock}$= 32 Km s$^{-1}$) and study the chemistry in the immediate and 
excursion region. The resulting abundances form the input for the shock 
at the next distance. The different parameters describing the shock structure 
at different radial distances from the star is given in Table 4. 

\subsection{Model results and discussion}

The molecular abundances of few selected species relative to the total
gas number density are given in Table~\ref{tbl:abundances} for various
shock strengths.  We have considered shock chemistry at gas layers
very close to the star.  The abundances at $2R_{\star}$ may not be the
exact values in the outflow because of uncertainties in the dust
formation radius which halts the shock chemistry, but are indicative.
One can see that in addition to molecular hydrogen certain other
molecules like CO, H$_2$O and N$_2$ are dominant, confirming that they
are the parent molecules.  But, to come to the point of our interest,
the shock chemistry is responsible for formation of species like HCN
along with CO$_2$ and CS close to the star.  The prediction of HCN at
the distance and environment described is in excellent agreement with
the observational result of Section~\ref{section:observations}.
Moreover, the theoretical value obtained for the HCN abundance is in
agreement with that derived from millimeter line observations in the
outer wind.

The chemical processes responsible for the formation of HCN involves reactions 
with cyanogen - an observation which was earlier found to be the case
in O-rich IK Tau (DCW99).  HCN is formed by the reaction 
\begin{equation}
\hbox{CN} + \hbox{H$_2$} \rightarrow \hbox{HCN} + \hbox{H}
\end{equation}
CN acts as an intermediary in the formation of HCN and CS and is destroyed 
quickly by atomic hydrogen. The reaction is sensitive to temperature and 
is very fast in the gas excursions. On the basis of the chemistry we can 
claim that HCN is a direct result of the shock chemistry in the inner 
wind region that we have considered here and travels as a parent species 
through the envelope unaltered until it reaches the photo-dissociation 
regions in the outer wind. 
The choice of the ratio C/O=0.95 is arbitrary, but with different values 
of C/O equal or slightly greater than 1.0, though the T.E. abundance values 
differ by a factor of 100, yet the resulting HCN abundance at 2R$_*$ remains 
within 5 percent of the quoted number in Table 5.

\section{Conclusions}
\label{section:conclusions}

The detection of HCN in high excitation states clearly demonstrates
that it is formed in the inner part of the envelope of $\chi$ Cyg,
within $\sol 20 R_{\star}$.  This is strongly supported by detailed
chemical modelling of the atmosphere.  The earlier claim of HCN along
with other carbon bearing molecules being formed in the innermost
region of the envelope of O -rich star like IK Tau (DCW99) seems to be
true in the case of S type star $\chi$ Cyg as well. This is the first
observational verification of the theoretical claim that shock
chemistry plays an active role in defining the chemical composition of
the inner wind regions of Miras.

\begin{acknowledgements}

The JCMT is operated by the Joint
Astronomy Centre on behalf of the PPARC, the Netherlands Organisation
for Scientific Research, and the National Research Council of Canada.
Our thanks to the JCMT staff who carried out observations on our behalf.
DD wants to thank A. J. Markwick for his help with the thermal equlibrium 
abundance calculation code. 

\end{acknowledgements}


\end{document}